\documentclass[12pt]{article}

\newcommand{\ot}{\otimes}
\newcommand{\be}{\begin{equation}} \newcommand{\ee}{\end{equation}}
\newcommand{\ba}{\begin{eqnarray}} \newcommand{\ea}{\end{eqnarray}}

\def\nn{\nonumber} 
\def\be{\begin{equation}}
\def\ee{\end{equation}}
\def\beq{\begin{equation}}
\def\eeq{\end{equation}}
\def\bea{\begin{eqnarray}}
\def\eea{\end{eqnarray}}
\def\ve{\vec{\epsilon}}
\def\si{\sigma}

\def\w{\vec{w}}
\def\u{\vec{u}}
\def\v{\vec{v}}
\def\n{\vec{n}}
\def\k{\vec{k}}

\def\T{\mathcal{T}}

\def\p{\partial}

\def\[{\{}

\def\]{\}}

\def\six{\sigma^x}
\def\siy{\sigma^y}
\def\siz{\sigma^z}

\begin{document}

\begin{center}
{\bf \Large Ladder Operators for Integrable One-Dimensional Lattice
Models}\\ \vspace{0.5cm}
M C Takizawa \footnote{mct@maths.uq.edu.au} and J R Links \\ \vspace{0.5cm}
Department of Mathematics, University of Queensland, Brisbane, QLD
4072, Australia
\end{center}

\begin{abstract}
A generalised ladder operator is used to construct the conserved operators
for any one-dimensional lattice model derived from the Yang-Baxter equation.  As an example, the
low order conserved operators for the $XYh$ model are calculated
explicitly.
\end{abstract}

\section{Introduction}
The method for constructing integrable one-dimensional lattice models from
solutions of the Yang-Baxter equation is well known (eg. see
\cite{faddeev}).  In principle, the conserved operators can be obtained by
series expansion of the family of commuting transfer matrices.  A more
practical approach is to use the ladder operator which permits a recursive
method through repeated commutators to obtain the conserved operators.

For models where the solution of the Yang-Baxter equation has the
difference property, it has been established \cite{thacker,sogo,tetelman} 
that the ladder operator is a lattice analogue of the boost operator for 
Lorentz invariant systems. Recently it has been shown that for the Hubbard 
model, which is not Lorentz invariant in the continuum limit as a consequence 
of spin-charge separation, and is reflected by the fact that the solution of 
the Yang-Baxter equation does not have the difference property, the ladder
operator still exists \cite{links}.

The present work extends \cite{links} to develop a general theory for the
construction of the ladder operator for any integrable system obtained
through the Yang-Baxter equation. The theory will be applied to analyse
the conservation laws for the XY model in a transverse magnetic field.

%The ladder operator method of calculating conserved currents has been
%applied successfully to the Heisenberg model
%\cite{thacker}-\cite{tetelman}.
%In these cases, the ladder operator is applicable only to models
%where the solution of the Yang-Baxter equation (YBE) has the difference
%property.
%Links et al. \cite{links}, however, constructed a general ladder operator
%$B$
%which can be applied to \emph{any} model that can be derived from a
%solution of the YBE.

\section{Integrable Lattice Models using the Quantum Inverse Scattering
Method}
%The starting point for the construction of integrable lattice models
%through the QISM is an invertible solution of the Yang-Baxter equation
%on a three-fold tensor product vector space $V\otimes V\ot V$
We begin with a vector-dependent solution of the Yang-Baxter equation
$$
%\be
R_{12}(\u,\v)R_{13}(\u,\w)R_{23}(\v,\w)=R_{23}(\v,\w)R_{13}(\u,\w)R_{12}(\u,\v)
%.\label{ybe}
$$
%\ee
where $\u$, $\v$ and $\w$ are $m$-component vectors. Throughout, we assume
the regularity property $R(\u,\u)=P$.
Define a set of $m$ local Hamiltonians
$$h_l\[i\]=P.\left.\frac{\p R_{l(l+1)}(\u,\v)}{\p u_i}\right|_{\u=\v},
~~i=1,..,m$$
with the corresponding global Hamiltonians acting on a one-dimensional
lattice of length $L$ given by
$$H\[i\]=\sum_{l=0}^{L-1} h_l\[i\]. $$
Throughout, periodic boundary conditions are assumed on all
summations which are evaluated over the length of the lattice.
Note it is implicit that all the
operators $h\[i\]$ are in fact functions of $\v$.

The transfer matrix is constructed through 
$$
%\be
T(\u,\v)={\rm
tr}_a\left(R_{a(L-1)}(\u,\v)...R_{a1}(\u,\v)R_{a0}(\u,\v)\right) 
%\ee
$$
where $a$ refers to the auxiliary space, which by the standard argument gives 
rise to a commutative family in the first variable; i.e.
 \bea
 &&[T(\u,\v),\,T(\w,\,\v)]=0,~~~\forall\, \u,\w. \label{ctm} 
 \eea
It can also be easily verified that
\be
[H\[i\],\,T(\u,\v)]=0,~~~\forall\,  \u. \label{int} \ee

It is convenient, however, to define the conserved operators as
$$
%\be
t\[\n\] = \left[ \frac{\p^{n_1+...+n_m}}{\p u_1^{n_1}...\p u_m^{n_m}}\ln
T(\u,\v) \right]
%\ee
$$
where they appear in the series expansion
\be
\ln T(\u,\v)=\sum_{\n} \frac {(u_1-v_1)^{n_1}...
(u_m-v_m)^{n_m}}{n_1!...n_m!}\,t\[\n\]. 
\label{expan} 
\ee
Thus it follows from (\ref{ctm}) that
$$[
t\[\n\],\, t\[\k\]]=0,~~~\forall\, \n,\k $$
and moreover from (\ref{int})
$$[H\[i\],\, t\[\n\]]=0,~~~\forall\, i, \n. $$
Note that $\n$ is an $m$-component vector with non-negative integer
entries.  Introducing the notation $\{\ve_i\}_{i=1}^m$ for
the basis of the $m$-dimensional vector space, we can write
$$\n=\sum_{i=1}^m n_i\ve_i.$$

\section{Recursion Formula for Calculating the Conserved Operators}
For each of the index labels $i$ we define a ladder operator
$$ B\[i\]=\sum_{l=0}^{L-1}  l  h_l\[i\]  $$
with the coefficients $l$ taken from the set of integers modulo $L$.
For any function $\phi$ admitting a Taylor's series expansion we have
$$[B\[i\],\,\phi(\T)]=\T.H\[i\].\phi^\prime(\T)$$
where $\T = T(\u,\u)$ and $\phi^\prime$ denotes the derivative of $\phi$.
Choosing $\phi$ to be the logarithm now gives
$$[B\[i\],\, \ln \T]= H\[i\].$$

%From the Sutherland relations (\ref{se}) it is deduced that
It can be shown that
\be
[B\[i\],\,T(\u,\v)]=-\frac{\p T(\u,\v)}{\p v_i}. \label{boost} \ee
As a result we obtain  the following recursion formula from (\ref{boost})
and the expansion (\ref{expan})
\be
t\[\n+\ve_i\]=[B\[i\], \, t\[\n\]] +\frac{ \p t\[\n\]}{\p v_i}
\label{recursion} \ee
The first few terms in (\ref{expan}) can be identified immediately
\be
t\[\vec 0 \]=\ln \T, ~~ ~~ ~~ t\[\ve_i\]=H\[i\]. \nn 
\ee
In principle, through repeated use of (\ref{boost}) expressions for all
the operators $t\[\n\]$ may be obtained. 

Applying the recursion (\ref{recursion}), the second order conserved currents 
can be obtained by the following formula:
\bea 
t\[\ve_i+\ve_j\]&=&  
 \frac12\sum_l\left[h_l\[j\],\,h_{l-1}\[i\]\right] 
+\frac12\sum_l[h_l\[i\],\,h_{l-1}\[j\]] \nn\\
&& ~~ +\frac12\frac{\p H\[j\]}{\p v_i}
+\frac12\frac{\p H\[i\]}{\p v_j}. 
\label{thingy}
\eea

\section{The XYh Model}
The $XY$ model in a transverse magnetic field has the following
Hamiltonian:
$$H = \sum_{i=1}^N (J_x \sigma_i^x \sigma_{i+1}^x + J_y \sigma_i^y 
\sigma_{i+1}^y + h \sigma_i^z) ~~ ~~ ~~ ~~ ~~ J_x, J_y, h \mbox{ const.}$$
This model is known to be integrable \cite{krinsky}.
Barouch and Fuchssteiner \cite{barouch}, Araki \cite{araki}, and Grabowski 
and Mathieu \cite{grabowski} have explicitly calculated the low order 
conserved operators.  These results have been reproduced using the
generalised ladder operator method.

\subsection{R Matrix of the XYh Model}
Bazhanov and Stroganov \cite{bazhanov} constructed an elliptic 
parametrization for the Boltzmann vertex weights of the $XYh$ model.   
In this parametrization, the weights are meromorphic functions of 3 complex 
variables, $\vec{u} = (u_1, u_2)$ and $\vec{v} = (v_1, v_2)$, where only the 
first vector entry contains the difference property.

The $R$ matrix is
$$R(\u,\,\v)
=\pmatrix{R^{11}_{11}&0&0&R^{11}_{22} \cr
0&R^{12}_{12}&R_{21}^{12}&0 \cr
0&R_{12}^{21}&R_{21}^{21}&0 \cr
R^{22}_{11}&0&0&R^{22}_{22} } $$
with
\bea
R^{22}_{22} & = & \rho (1-e(u_1-v_1)e(u_2)e(v_2))   ~~ ~~ ~~ ~~ ~~ 
R^{11}_{11}  = \rho (e(u_1-v_1)-e(u_2)e(v_2)) \nn \\
R^{21}_{21} & = & \rho (e(u_2)-e(u_1-v_1)e(v_2)) ~~  ~~ ~~ ~~ ~~   ~~
R^{12}_{12} = \rho (e(v_2)-e(u_1-v_1)e(u_2)) \nn \\
R^{12}_{21} & = & R^{21}_{12} =
\frac{\rho\sqrt{e(u_2)s(u_2)}\sqrt{e(v_2)s(v_2)}(1-e(u_1-v_1))}{s
\left( \frac{u_1-v_1}{2} \right) } \nn \\
R^{11}_{22} & = & R^{22}_{11} =
-ik\rho\sqrt{e(u_2)s(u_2)}\sqrt{e(v_2)s(v_2)}(1+e(u_1-v_1))
s\left( \frac{u_1-v_1}{2} \right) \nn
\eea
where $\rho$ is an arbitrary constant and $s$ and $e$ are the respective 
elliptic functions $sn$ and $(cn +i\, sn)$.  By imposing the condition 
$ R(\u,\,\u) = P$, we obtain the value of $\rho = \frac{1}{1-e^2(v_2)}$.

\subsection{Local Hamiltonians}
The local Hamiltonians $h_l\[1\]$ and $h_l\[2\]$ are given as follows:
\bea
h_l\[1\] 
&=&\rho [A(\si^x\ot\si^x)+ B(\si^y\ot\si^y)+ C(I\ot I)+ D(I\ot \si^z +
\si^z\ot I)]\nn \\
h_l\[2\] 
&=&\rho [E(I\ot I)+ F(\si^x\ot\si^y-\si^y\ot\si^x)] \nn
\eea
where $\si^x, \si^y, \si^z$ are the Pauli sigma matrices and
\bea
&& A = -\frac{1}{2}ie(v_2)(1+ks(v_2)), ~~ ~~ 
B= -\frac{1}{2}ie(v_2)(1-ks(v_2)), 
~~ ~~ ~~ k \mbox{ const.} \nn \\
&& C = s(v_2)e(v_2),  ~~ ~~ ~~ ~~ ~~ ~~ ~~ ~~ ~~ ~~ ~~ 
D = \frac{1}{2}ic(v_2)e(v_2), \nn \\
&& E = -id(v_2)e(v_2)^2,~~ ~~ ~~ ~~ ~~ ~~ ~~ ~~ 
F = \frac{1}{2}d(v_2)e(v_2). \nn
\eea
where $d$ and $c$ are the elliptic functions $dn$ and $cn$ respectively.

\subsection{Second Order Conserved Currents}
The second order conserved currents can be obtained by the formula
(\ref{thingy}):
\begin{eqnarray*}
t\[2\ve_1\] = t\[2\ve_2\] &=& \sum_l \left\{ \alpha (\six\ot\siz\ot\siy
- \siy\ot\siz\ot\six) \right.\\
&& \left. ~~~ + \beta (\six\ot\siy\ot I - \siy\ot\six\ot I) \right\}
~~~~~~ \alpha ,\beta\mbox{ const.}
\end{eqnarray*}
and
%\bea
%t\[\ve_1+\ve_2\]
%&=&\rho^2\sum_l\{P(\si^x\ot\si^z\ot\si^x)+ Q(\si^y\ot\si^z\ot\si^y)\nn\\
%&& ~~ ~~ +R(\si^x\ot\si^x\ot I)+S(\si^y\ot\si^y\ot I)\nn\\
%&& ~~ ~~ +T(\si^z\ot I\ot I)+U(I\ot I\ot I)\}\nn
%\eea
%where
%\bea
%&& P=2iAF,~~ ~~ ~~ ~~ ~~ ~~ ~~ ~~ ~~ ~~ ~~ ~~ ~~ ~~ ~~ ~~ ~~ ~~ ~~ ~~ ~~ ~~ 
%Q=2iBF,\nn\\
%&& R=\frac{1}{2}(\rho 'A+\rho A')(1-e_3^2)^2-2iDF, ~~
%S=\frac{1}{2}(\rho 'B+\rho B')(1-e_3^2)^2-2iDF,\nn\\
%&& T=(\rho 'D+\rho D')(1-e_3^2)^2, ~~ ~~ ~~ ~~ ~~ ~~ ~~ ~~
%U=\frac{1}{2}(\rho 'C+\rho C')(1-e_3^2)^2\nn
%\eea
%which simplifies to 
\bea
t\[\ve_1+\ve_2\]
&=&\sum_l\{\gamma(\si^x\ot\si^z\ot\si^x)+ 
\zeta(\si^y\ot\si^z\ot\si^y)\nn\\
&& ~~ ~~ +\eta (\si^x\ot\si^x\ot I + \si^y\ot\si^y\ot I)\nn\\
&& ~~ ~~ - (\gamma +\zeta )(\si^z\ot I\ot I)\} ~~ ~~ ~~ ~~ ~~ ~~ ~~ ~~ ~~
~~ ~~ ~~ ~~ ~~ ~~
\gamma, \zeta, \eta \mbox{ const.}\nn
\eea
in agreement with \cite{barouch}-\cite{grabowski}.

\end{document}